\pdfoutput=0
\documentclass[11pt]{article}
\usepackage{epsfig}
\usepackage{amsfonts}
\usepackage{amsmath}
\usepackage{bm,bbm}
\usepackage{cite}
\allowdisplaybreaks[1]

\newcommand{\la}[1]{\label{#1}}
\newcommand{\be}{\begin{equation}}
\newcommand{\ee}{\end{equation}}
\newcommand{\ba}{\begin{eqnarray}}
\newcommand{\ea}{\end{eqnarray}}
\newcommand{\rmi}[1]{{\mbox{\scriptsize #1}}}
\newcommand{\fig}{Fig.~}
\newcommand{\figs}{Figs.~}
\newcommand{\eq}{Eq.~}
\newcommand{\eqs}{Eqs.~}
\newcommand{\se}{Sec.~}

\newcommand{\nr}[1]{(\ref{#1})}
\newcommand{\tr}{{\rm Tr\,}}

\newcommand{\fr}[2]{{\frac{#1}{#2}\,}}
\newcommand{\msbar}{{\overline{\mbox{\rm MS}}}}

\renewcommand{\vec}[1]{{\bf #1}}

\renewcommand{\eq}{eq.~}
\renewcommand{\eqs}{eqs.~}
\renewcommand{\se}{sec.~}

\renewcommand{\fig}{fig.~}
\renewcommand{\figs}{figs.~}

\newcommand{\tinymsbar}{{\overline{\mbox{\tiny\rm{MS}}}}}
\newcommand{\Lambdamsbar}{{\Lambda_\tinymsbar}}

\newcommand{\T}{\rmii{$T$}}
\newcommand{\R}{\rmii{$R$}}
\newcommand{\RB}{\rmii{$R,\!B$}}
\newcommand{\RBtiny}{{\mbox{\tiny{$\!\scriptstyle{R,\!B}$}}}}
\newcommand{\mpl}{m^{ }_\rmii{Pl}}

\newcommand{\Nc}{N_{\rm c}}

\newcommand{\mE}{m_\rmii{E}}

\newcommand{\rmO}{{\mathcal{O}}}
\newcommand{\bmu}{\bar\mu}

\newcommand{\dA}{d_\rmii{A}^{ }}

\def\lsi{\raise0.3ex\hbox{$<$\kern-0.75em\raise-1.1ex\hbox{$\sim$}}}
\def\gsi{\raise0.3ex\hbox{$>$\kern-0.75em\raise-1.1ex\hbox{$\sim$}}}
\newcommand{\lsim}{\mathop{\lsi}}
\newcommand{\gsim}{\mathop{\gsi}}

\newcommand{\rmii}[1]{{\mbox{\tiny\rm{#1}}}}
\newcommand{\rmiii}[1]{{\mbox{\tiny{$\scriptstyle{\rm#1}$}}}}
\newcommand{\re}{\mathop{\mbox{Re}}}
\newcommand{\im}{\mathop{\mbox{Im}}}

\newcommand{\Tint}[1]{{\hbox{$\sum$}\!\!\!\!\!\!\!\int\,}_{\!\!\!\!\raise-0.9ex\hbox{$\scriptstyle{#1}$}}}
\newcommand{\Tinti}[1]{{{\Sigma}\!\!\!\!\raise0.3ex\hbox{$\int$}_\rmii{${#1}$}}}

\newcommand{\bi}{\begin{itemize}}
\newcommand{\ei}{\end{itemize}}
\newcommand{\hide}[1]{ }

\makeatletter \@addtoreset{equation}{section} \makeatother
\renewcommand{\theequation}{\arabic{section}.\arabic{equation}}
\makeatletter
\renewcommand\section{\@startsection {section}{1}{\z@}%
                                   {-5.5ex \@plus -1ex \@minus -.2ex}
                                   {2.3ex \@plus.2ex}%
                                   {\normalfont\large\bfseries}}
\renewcommand\subsection{\@startsection{subsection}{2}{\z@}%
                                     {-3.25ex\@plus -1ex \@minus -.2ex}%
                                     {1.5ex \@plus .2ex}%
                                     {\normalfont\normalsize\bfseries}}
\renewcommand\thesection {\@arabic\c@section}
\renewcommand\thesubsection   {\thesection.\@arabic\c@subsection}
\renewcommand{\@seccntformat}[1]{%
\csname the#1\endcsname.\hspace{1.0em}}
\makeatother

\begin{document}

\flushbottom

\begin{titlepage}

\begin{flushright}
June 2021
\end{flushright}
\begin{centering}
\vfill

{\Large{\bf
 Minimal warm inflation with complete medium response
}} 

\vspace{0.8cm}

M.~Laine 
 and 
S.~Procacci

\vspace{0.8cm}

{\em
AEC, 
Institute for Theoretical Physics, 
University of Bern, \\ 
Sidlerstrasse 5, CH-3012 Bern, Switzerland \\}

\vspace*{0.8cm}

\mbox{\bf Abstract}
 
\end{centering}

\vspace*{0.3cm}
 
\noindent
If a homogeneous field evolves within a medium, with the latter
gradually picking up a temperature, then the friction felt by the
field depends on how its evolution rate compares with medium time
scales. We suggest a framework which permits to
incorporate the contributions from all medium time scales. As an
example, we illustrate how warm axion inflation can be described by
inputting the retarded pseudoscalar 
correlator of a thermal Yang-Mills plasma. 
Adopting a semi-realistic model for the latter, 
and starting the evolution at almost vanishing temperature, 
we show how the system heats up and then enters
the ``weak'' or ``strong'' regime of warm inflation. 
Previous approximate treatments are scrutinized. 

\vfill


\end{titlepage}

\tableofcontents

%
\section{Introduction}
\la{se:intro}

A general empirical observation from interacting multiparticle systems is
that they tend to equilibrate fast, 
often after a handful of elementary scatterings. 
In the context of inflation, such a ``reheating'' is 
expected to take place after a period of exponential expansion
(cf.,\ e.g.,\ ref.~\cite{lozanov}). 
At the same time, the inflationary expansion itself is normally assumed
to be non-thermal~\cite{linde}. 
One reason is that a thermal mass generated for the inflaton
field would in general compromise the desired 
flatness of the potential. 

The problem with large thermal corrections can, however, 
be evaded with certain types
of interactions. In the present paper we are concerned with
the example of axion inflation~\cite{ai}, 
which at its late stage can be described by 
\be
 \mathcal{L} 
 \supset
 \frac{1}{2}
 \bigl( 
     \partial^\mu\varphi\, \partial_\mu\varphi - m^2 \varphi^2 
 \bigr) 
 - \frac{\varphi \,  \chi}{f^{ }_a}
 \;, 
 \quad
 \chi \equiv \frac{ 
 \epsilon^{\mu\nu\rho\sigma}_{ }
 g^2 F^{c}_{\mu\nu} F^{c}_{\rho\sigma}
 }{64\pi^2}
 \;, 
 \quad
 c \in \{ 1,..., \Nc^2 - 1  \} 
 \;, \la{L}
\ee
where $F^c_{\mu\nu}$ is the Yang-Mills field strength, $\Nc^{ }$ is 
the number of colours, and $f^{ }_a$ is the axion decay constant. 
The thermal mass generated by this interaction 
vanishes to all orders in perturbation theory.
It does not vanish non-perturbatively, however  
the value is very small at temperatures above 
the confinement scale (cf.,\ e.g.,\ refs.~\cite{chi1,chi2} 
and references therein).\footnote{%
 Axion inflation has been proposed to evade the steepness of 
 the potential from other considerations as well, restricting
 to an Abelian gauge field and studying its dynamics on 
 a classical level rather than letting it thermalize~\cite{axion1}
 (see, e.g.,\ ref.~\cite{axion3} for recent work
 and references). Thermalized cases with non-Abelian gauge
 fields have also been considered
 (cf.,\ e.g.,\ refs.~\cite{axT00,axT0,axT1,axT2,axT25,axT3} 
 and references therein), 
 however the thermal friction coefficient 
 did not have the form
 that is believed to dominate at high temperatures (see below).
 }

Recently, it has been stressed that the system in \eq\nr{L} can 
lead naturally~\cite{warm1,warm2,warm3,warm4} 
to warm inflation~\cite{warm_old,warm_rev}.\footnote{%
 Only ref.~\cite{warm2} considered exactly the setup
 of \eq\nr{L}, and then not for inflation but rather 
 for early dark energy. 
 Ref.~\cite{warm1} considered ``hybrid inflation'', 
 implying that \eq\nr{L} is supplemented
 by a constant during the inflationary period, which 
 rapidly switches off as inflation ends.  
 Ref.~\cite{warm3} had an exponential potential, 
 ref.~\cite{warm4} the periodic 
 $\sim \Lambda^4 [1 + \cos(\varphi/f^{ })]$, which 
 is equivalent to \eq\nr{L} around the end when $\varphi\approx \pi f^{ }$.
 } 
Apart from the absence of a thermal mass correction, an essential 
ingredient for this is a thermal friction coefficient, 
which slows down the inflaton motion and keeps the Yang-Mills field 
at a finite temperature. The additional thermal friction may permit for
successful inflation in spite of largish slow-roll parameters, 
thereby possibly helping 
to evade so-called swampland concerns~\cite{swampland1,swampland2}. 

In refs.~\cite{warm1,warm2,warm3,warm4}, the value of the thermal friction 
coefficient was adopted from the domain of small frequencies, 
$\omega \lsim \alpha^2 T$
(here $\alpha \equiv g^2/(4\pi)$ is the Yang-Mills coupling),
where it is related 
to the so-called Chern-Simons diffusion rate~\cite{mms}, 
known from numerical simulations at temperatures
far above the confinement scale~\cite{sphaleron}. However, 
at early stages of warm inflation, when the 
temperature is low, the rate of change of $\varphi$ 
is relatively speaking larger, 
$\omega \sim \pi T$ or even $\omega \gg \pi T$. 
The purpose of the current study is to present a framework
permitting to study all of these domains. This should include 
an initial state with $T \approx 0$ and, if appropriate, a crossover
from Hubble friction to thermal friction dominated dynamics.   

%
\section{Rates and frequency scales}
\la{se:scales}

The system described by \eq\nr{L} is sensitive to a number of scales: 
$m, f^{ }_a$, as well as the confinement scale of the Yang-Mills
plasma, $\Lambdamsbar$. The thermal environment brings in the
temperature, $T$; cosmology brings in the Planck mass, 
$\mpl \approx 1.221 \times 10^{19}$~GeV; 
and in addition
we need to insert an initial value for the inflaton field, $\varphi(0)$.
The initial $\dot\varphi(0)$ is chosen so that we are 
close to a slow-roll regime. For simplicity 
we assume that after thermalization,  
$T \gg \Lambdamsbar$ and $\alpha \ll 1$, however 
our theoretical ingredients also apply in a strongly coupled regime,
even if the practical implementation would be much harder there.

A key quantity for warm inflation is a thermal friction coefficient, 
denoted traditionally by $\Upsilon$. Letting $H$ be the Hubble rate, 
a ``strong regime'' of warm inflation refers to 
$Q \equiv \Upsilon/(3 H) \gg 1$, a ``weak regime'' to $Q \lsim 1$.
The total friction felt by $\varphi$ is $3H(1+Q)$. 

In chaotic inflation~\cite{chaotic}, 
where $f^{ }_a \to \infty$ and $Q = 0$, 
a sufficient number of $e$-folds is obtained
if $\varphi(0) \;\gsim\; (\mbox{a few})\times \mpl$, and the
COBE normalization of temperature fluctuations 
requires $m \ll \mpl$. It is argued that in
the strong regime of minimal warm inflation, it is sufficient to have 
$\varphi(0) \;\lsim\; \mpl$, as the additional friction
slows down the evolution of the scalar field~\cite{warm1}. 
In the weak regime, thermal effects play a role only
towards the end of the inflationary period, nevertheless 
they have been argued to shift the spectral 
tilt $n^{ }_s$ in a favourable direction~\cite{warm4}. 

As the operator $\varphi\,\chi / f^{ }_a$ added in \eq\nr{L} is
non-renormalizable, certain constraints need to be satisfied
in order for the framework to be self-consistent. If we consider on-shell 
wave modes with $\omega^2\sim m^2$, then there is 
an ultraviolet divergence to the self-energy,  
of magnitude $\sim \alpha^2 m^4/f_a^2$
(cf.\ \eq\nr{C_R_vac}).
This should be a small correction, so we require
\be
 \alpha^2 m^2 \ll f_a^2
 \;, 
 \la{consistency}
\ee
implying that $m$ can be at most modestly larger than $f^{ }_a$.\footnote{%
 For completeness, we recall that in a strongly coupled system, 
 such as chiral perturbation theory, the requirement 
 $m \ll f^{ }_a / \alpha$ is replaced by 
 $m^{ }_\pi \ll 4\pi f^{ }_\pi$.
 }
On the other hand, assuming a cold initial state, the effects from
$\varphi\chi/f_a$ turn out to be insignificant far from the regime
$\alpha^2m^2\sim f_a^2$ . 
Once we approach this regime, \eq\nr{Ups_vac} implies
that the width of the scalar field is of the same order as its mass. 
Then the inflaton decays very fast, 
generating thereby an ensemble of Yang-Mills
bosons, which can subsequently thermalize.

As far as $\Upsilon$ is concerned, the damping rate of the scalar
field depends strongly on the frequency scale at which it is evolving. 
In the limit of vacuum decays we have (cf.\ \eq\nr{C_R_vac}) 
\be
 \Upsilon^{ }_\rmiii{UV} \sim \frac{\alpha^2 m^3}{f_a^2}
 \;, \la{Ups_vac}
\ee
whereas for slow thermal processes 
the rate is (cf.\ \eq\nr{rho_T_ir}) 
\be
 \Upsilon^{ }_{\rmiii{IR}} \sim \frac{\alpha^5 T^3}{f_a^2}
 \;. \la{Ups_T}
\ee
The comparison of these rates to $H$ determines whether we
are in the weak or strong regime. 

A further quantity playing a key role is what may
be referred to as the thermalization rate of the Yang-Mills plasma, 
\be
 \Delta \sim \alpha^2 T
 \;. \la{Delta_appro}
\ee
Only if the system is probed with a frequency 
$\omega \sim m \lsim \Delta$, 
does \eq\nr{Ups_T} represent the correct interaction rate. 
For $\omega \gg \Delta$, i.e.\ particularly at the beginning, 
we should rather use \eq\nr{Ups_vac}.
This setup guarantees that thermal physics plays a substantial role 
only in a domain where the assumption of a sufficient thermalization 
time is self-consistent.

Now, the value of $T$, and subsequently those of $\Delta$ and 
$\Upsilon$, are determined by the
dynamics of the solution, i.e.\ by the parameters 
$m$, $H$, and also
by the values of $\Delta$ and $\Upsilon$ themselves. 
This non-linear dependence implies
that we cannot fix $\Upsilon$ in advance, but
that all eventualities need to be accounted for by the basic
equations. This makes our setup, introduced in \se\ref{se:eom_scalar}, 
more complicated than those in  
refs.~\cite{warm1,warm2,warm3,warm4}, 
which adopted \eq\nr{Ups_T} for all frequencies.

%
\section{Scalar field equation of motion}
\la{se:eom_scalar}

For a quantitative study, we need to establish
the equation of motion satisfied by $\varphi$ in the
presence of a heat bath.\footnote{%
 If $\varphi$ evolves instead within a non-equilibrium ensemble,
 there may be additional contributions to the friction coefficient
 (cf.,\ e.g.,\ ref.~\cite{axion1}).
 This could be important particularly if the ensemble
 consists of Abelian gauge fields, which are less likely to thermalize,
 and generate no sphaleron contribution from their thermal modes. 
 }
We start by recalling the general linear response
argument for how the dynamics of a weakly coupled field and a heat bath
are connected to each other. We first proceed ``blindly'', assuming that
all integrals are well-defined; subsequently, short-distance
singularities that are inevitable in quantum field theory are 
incorporated. The logic bears certain resemblances e.g.\ to the 
classic ref.~\cite{mms} even if we go beyond it in many ways. 

According to the equivalence principle, we are free to choose the frame
in which to derive the equation of motion. For thermal field theory and
statistical physics, it is convenient to operate in the medium rest frame,
such that its four-velocity is $u = (1,\vec{0})$ and temperature 
appears in its text-book form. 
It is also helpful that
all parameters are, to a good approximation, time-independent. For these 
reasons we first consider a locally Minkowskian frame. Subsequently
the equation of motion is written in a covariant form, 
and we can then easily incorporate an expanding 
Friedmann-Lema\^itre-Robertson-Walker metric. 

Consider a theory defined by 
\be
 \mathcal{L} = \fr12
 \bigl(
  \partial^{\mu}\varphi\, \partial_{\mu}\varphi 
  - m^2 \varphi^2  
 \bigr)
 - \varphi\, J + \mathcal{L}^{ }_\rmi{bath}
 \;, \la{L_2}
\ee
where $J$ is a gauge-invariant composite operator 
($J = \chi/f^{ }_a$ in the case of \eq\nr{L}) and 
$\mathcal{L}^{ }_\rmi{bath}$ is the Lagrangian for 
the heat bath degrees of freedom. If we assume that 
$\varphi$ is spatially constant and evolves slowly, it should
satisfy a classical equation of motion derived from \eq\nr{L_2}, 
\be
 \ddot{\varphi} + m^2 \varphi + 
 \langle J(t) \rangle = 0 
 \;. 
\ee
The last term is important because the average value 
of the composite 
operator depends on the slowly evolving
value of the $\varphi$-background. 

To determine $ \langle J(t) \rangle $, we can inspect how 
the density matrix of the heat bath evolves with time
in the presence of a $\varphi$-background. Let us
write the heat bath Hamiltonian as 
$\hat{H} = \hat{H}^{ }_\rmi{bath} + \varphi\, \hat{J}$, 
and assume that the initial density matrix $\hat{\rho}(0)$ is an equilibrium
state, i.e.\ $[\hat{H}^{ }_\rmi{bath},\hat{\rho}(0)] = 0$. We now solve the
evolution equation 
$
 i \partial^{ }_t \hat{\rho}(t) = 
 [ \hat{H}(t),\hat{\rho}(t) ] 
$
to first order, 
\be
 \hat{\rho}(t) = \hat{\rho}(0) - i 
 \int_0^t \! {\rm d}t' \, 
 [ \hat{H}(t'),\hat{\rho}(0) ] + \ldots
 \;, 
\ee
and note that here $\hat{H}(t')$ can be replaced 
with $\varphi(t') \hat{J}(t')$, 
because $\hat{H}^{ }_\rmi{bath}$ commutes with 
$\hat{\rho}(0)$. Inserting
this solution we can compute 
\be
 \langle \hat{J}(t) \rangle 
 \; \equiv \; 
 \tr [\hat{\rho}(t) \hat{J}(t)]
 \; = \; 
 \langle \hat{J}(0) \rangle^{ }_0 
 - \int_0^t \! {\rm d}t' \, \varphi(t') \, C^{ }_\R(t-t') + 
 \rmO(J^3) 
 \;, \la{Jt}
\ee
where the retarded correlator is defined as 
\be
 C^{ }_\R(t - t')
 \; \equiv \;  
 \theta(t-t')\, \bigl\langle
 i 
 \, 
 \bigl[ 
  \hat{J}(t) , \hat{J}(t') 
 \bigr]
 \bigr\rangle^{ }_0 
 \;, \la{C_R_def} 
\ee
and the expectation value $\langle .... \rangle^{ }_0$ 
is taken with respect to $\hat{\rho}(0)$. 
As the time evolution of the Heisenberg operator 
$\hat{J}(t)$ is given by $\hat{H}^{ }_\rmi{bath}$, 
we have used
$
  \langle \hat{J}(t) \rangle^{ }_0 =
  \langle \hat{J}(0) \rangle^{ }_0
$
in \eq\nr{Jt}.
Assuming $ \langle \hat{J}(0) \rangle^{ }_0 = 0$ because 
$\hat{J}$ is odd under discrete symmetries, 
and re-expressing the time integration domain,  we end up with
\be
 \ddot{\varphi}(t) + m^2 \varphi(t) - 
 \int_0^\infty \! {\rm d}t' \, 
 C^{ }_\R(t-t') \, \varphi(t') 
 = 0
 \;, \quad t \ge 0
 \;. \la{eom}
\ee

At this point, we note from \eq\nr{C_R_vac} that the 
Fourier transform of $C^{ }_\R$ grows like $\sim \omega^4$ at large
frequencies (with a logarithmically divergent coefficient). This implies
that $C^{ }_\R(t-t')$ diverges like $\sim 1/(t-t')^5$ at short times. 
Then neither 
the convolution integral in \eq\nr{eom}, nor the Fourier transform of 
$C^{ }_\R$, are well-defined. 

Let us assume that, as is usual in quantum field theory, 
the divergences can be taken care of by introducing some 
regularization and subsequently cancelling them by local counterterms. 
Given the degree of divergence, this requires that we
modify \eq\nr{eom} into
\be
 \delta Z\, \varphi^{(4)}_{ }(t) + 
 \ddot{\varphi}(t) + m^2 \varphi(t) - 
 \int_0^\infty \! {\rm d}t' \, 
 C^{ }_\RB(t-t') \, \varphi(t') 
 = 0
 \;, \la{eom2}
\ee
where $C^{ }_\RB$ now stands for a ``bare'' correlator. 
The regularization prescription is left implicit. 

We now analyse \eq\nr{eom2} through 1-sided Fourier transforms
(or Laplace transforms). Defining
\be
 \tilde\varphi(\omega) \, \equiv \, 
 \int_0^{\infty} \! {\rm d}t\, e^{i \omega t} \, \varphi(t)
 \;, \la{transform}
\ee
and assuming that $\varphi(t)$ grows at most power-like at large $t$
(physically, its absolute value decreases at large $t$), 
we note that $\varphi$ is analytic in the upper half
of the complex plane ($\im\omega > 0$). 
The inverse transform is a 2-sided one, 
\be
 \varphi(t) \theta(t) 
 = 
 \int_{-\infty}^{\infty}
 \! \frac{{\rm d}\omega}{2\pi} \, 
 e^{-i \omega t} \, \tilde\varphi(\omega)
 \;. \la{inverse}
\ee
Similarly, 
$ C^{ }_\R(\omega) = \int_0^\infty \! {\rm d}t \, e^{i\omega t}
  C^{ }_\R(t)
$ 
is analytic in the upper half-plane.

Inserting the Fourier transforms and 
making use of the convolution theorem,\footnote{
 Or, concretely, 
 multiplying \eq\nr{eom2} by $e^{i(\omega + i 0^+)t}$,  
 integrating over $t\ge 0$, 
 evaluating 
 $ 
 \int_{-\infty}^{\infty} \! \frac{{\rm d}\omega'}{2\pi i}
 \frac{C^{ }_\RBtiny(\omega') \tilde\varphi(\omega')}
      {\omega' - \omega - i 0^+}
 $
 with the Cauchy theorem, 
 again assuming the presence of regularization, 
 which cuts off $C^{ }_\RB$ at large $|\omega'|$.}
we find
\be
 \bigl[ \, 
 -\delta Z \, \omega^4 
 + \omega^2  - m^2 
 +  
 C^{ }_\RB(\omega)
 \, \bigr]\, \tilde\varphi(\omega)
 \; = \;  
 \mathcal{G}[\omega,\varphi^{(n)}_{ }(0)]
 \;, \la{eomF2}
\ee
where $\mathcal{G}$ contains all terms related to initial conditions, 
\be
 \mathcal{G}[\omega,\varphi^{(n)}_{ }(0)] 
 \; = \; 
 \delta Z \, 
 \bigl[ -\varphi^{(3)}_{ }(0) + i \omega \ddot{\varphi}(0)
        + \omega^2 \dot{\varphi}(0) 
        - i \omega^3  \varphi(0) \bigr]
 -
   \dot{\varphi}(0)
 + 
  i\omega 
 \varphi(0)
 \;. \la{G}
\ee
Eq.~\nr{eomF2} is immediately solved and transformed back as  
\be
 \varphi(t)\theta(t)
 = 
 \int_{-\infty}^{\infty}
 \! \frac{{\rm d}\omega}{2\pi} \, 
 \frac{e^{-i \omega t}}
 {-\delta Z\,\omega^4 + \omega^2 - m^2 + C^{ }_\RB(\omega)
 }
 \, 
 \mathcal{G}[\omega,\varphi^{(n)}_{ }(0)]
 \;. \la{eomF3} 
\ee
We assume in the following that $\delta Z$ cancels the divergences
of $C^{ }_\RB(\omega)$, and therefore omit both $\delta Z$
and the divergences from $C^{ }_\RB$, 
returning to the notation $C^{ }_\R$. 
It should also be mentioned that the strange-looking 
\eq\nr{G} will not be needed in practice, however it illustrates
that initial conditions are subject to renormalization as well. 

Now, the idea is to consider a ``macroscopic'' $t > 0$ and then to 
deform the integration contour in 
\eq\nr{eomF3} into the lower half-plane. The deeper we can 
go, the faster is the exponential fall-off of the solution. 
Hence, the slowest dynamics of $\varphi$ can be identified
by searching for the singularities closest to the real axis. 
We note in passing that 
$
  \mathcal{G}[\omega,\varphi^{(n)}_{ }(0)]
$
has zeros, however their locations depend on initial conditions, 
and therefore cannot coincide in general with the roots of the denominator, 
which are independent of initial conditions. 

An alert reader may worry about the need to know $C^{ }_\R$
in the lower half-plane. Indeed many text-book relations, such as the 
spectral representation 
\be
 C^{ }_\R(\omega + i 0^+_{ }) = 
 \int_{-\infty}^{\infty}
 \! \frac{{\rm d}\omega'}{\pi} \frac{\rho(\omega')}{\omega' - \omega - i 0^+}
 \;, \la{spectral}
\ee
where $\rho(\omega) \equiv \im C^{ }_\R(\omega + i 0^+)$ is called 
the spectral function, concern the side of the upper half-plane.
Furthermore, $0^+$ cannot be made negative, 
since $1/(\omega' - z)$ is discontinuous across the real axis:
$
  1/(\omega'-\omega-i 0^+_{ })
 - 
  1/(\omega'-\omega+i 0^+_{ }) 
 = 2\pi i \delta(\omega' - \omega)
$. 
That said, 
$C^{ }_\R$ is still defined in the lower half-plane; 
it is just not analytic there, but must have singularities.\footnote{%
 A possible
 way to determine $C^{ }_\R$ is to solve the Cauchy-Riemann differential
 equations, taking $ C^{ }_\R(\omega + i 0^+) $ as the initial condition. 
 This system can be rephrased as a 2-dimensional Laplace equation.  
 It is known that the solution of the Laplace equation with 
 Cauchy boundary conditions 
 is unstable, reflecting the singularities.}
In the following we assume that the singularity 
structure of $C^{ }_\R$ is known in the lower half-plane, 
and return in \se\ref{se:C_R} to examples 
for how it could look like (there may be poles and cuts).  

With a given $C^{ }_\R$, let us search for the roots of the denominator.
Concretely, we inspect 
\be
 \ddot{\varphi} + \Upsilon \dot{\varphi} + m_\T^2 \varphi
 = 
 - \int_{-\infty}^{\infty}
 \! \frac{{\rm d}\omega}{2\pi} \, 
 \frac{\omega^2 + \Upsilon i\omega - m_\T^2 }
 {\omega^2 - m^2 + C^{ }_\R(\omega)
 }
 \, 
 e^{-i \omega t}
 \, 
 \mathcal{G}[\omega,\varphi^{(n)}_{ }(0)]
 \;. \la{eomF4} 
\ee
The parameters $\Upsilon$ and $m_\T^2$ are to be chosen 
such that the leading singularities are lifted.
We note that the ``literal'' initial conditions encoded in 
$\mathcal{G}[\omega,\varphi^{(n)}_{ }(0)]$ 
play no role, as the equation of motion obtained
by cancelling the poles is valid only at large times. 
 
Given that $C^{ }_\R$ is generated by a non-renormalizable
interaction,
the whole setup is consistent only to the extent that
$C^{ }_\R$ is treated as a small correction.\footnote{%
 It should be possible to construct renormalizable models
 of warm inflation as well (cf.,\ e.g., ref.~\cite{eft}).
 Our discussion applies to 
 these cases provided that the inflaton is coupled weakly
 to the plasma degrees of freedom. 
 }  
In this 
situation, we can solve for the roots iteratively. At tree-level,
the roots are at $\omega = \pm m$. 
The symmetries 
$\re C^{ }_\R(-m) = \re C^{ }_\R(m)$ and 
$\im C^{ }_\R(-m) = - \im C^{ }_\R(m)$
imply that we can restrict to the root at $\omega = + m$.
Denoting 
\be
 C^{ }_\R(m) = \re C^{ }_\R + i \im C^{ }_\R
 \;, 
\ee
the desired parameters are given by 
\be
 \Upsilon \approx \frac{\im C^{ }_\R}{m}
 \;, \quad
 m_\T^2 \approx m^2 - \re C^{ }_\R 
 \;.
 \la{eff_params} 
\ee

In the limit of slow evolution 
($m  \ll \alpha^2 T$), 
the real part of $C^{ }_\R$ gives the correction
\be 
 \delta m^2_{\rmiii{IR}} 
 \; \equiv \; 
 - \re C^{ }_\R(0)
 \;. 
 \la{dm2}
\ee
The negative sign appears because we are in Minkowskian spacetime; 
a Wick rotation to Euclidean spacetime inserts an imaginary unit to $\chi$, 
and then \eq\nr{dm2} corresponds to the topological susceptibility
that is being measured in lattice simulations~\cite{chi1}. 
For $m  \ll \alpha^2 T$, 
the imaginary part of $C^{ }_\R$
produces a thermal friction coefficient
in accordance with refs.~\cite{moduli1,basics}, 
\be
 \Upsilon^{ }_{\rmiii{IR}} 
 \; \equiv \;  
 \lim_{\omega\to 0} (-i) C'_\R(\omega) 
 \;.
 \la{slow_Gamma}
\ee
More generally, $\re C^{ }_\R$ and $\im C^{ }_\R$ are evaluated at $m > 0$, 
and can have different values.

Having derived the basic equation in locally Minkowskian spacetime, we can 
promote the result originating from \eq\nr{eomF4} 
to a general coordinate system,  
\be
 {\varphi^{;\mu}_{ }}^{ }_{;\mu} 
 + \Upsilon u^\mu_{ }\varphi^{ }_{;\mu} + m_\T^2 \varphi
 \; \simeq \; 
 0 
 \;. \la{eom_covariant} 
\ee
Restricting finally to a homogeneous field, yields our final scalar 
equation of motion, 
\be
  \ddot{\varphi} + (3 H + \Upsilon)\dot{\varphi} + m_\T^2 \varphi
 \;\simeq\;
  0
 \;. \la{eomF5} 
\ee
The form agrees with refs.~\cite{warm1,warm2,warm3,warm4}, 
however $\Upsilon$ obtained from \eq\nr{eff_params}
interpolates between \eqs\nr{Ups_vac} and \nr{Ups_T}, and 
$m_\T^2$ may contain a non-trivial correction as well. 

%
\section{Heat bath equation of motion}
\la{se:eom_bath}

Usually, the mass parameter $m^2$ is assumed to be temperature-independent, 
and we have argued below \eq\nr{L} 
that this is the case in the present system to 
a good approximation. However, taken literally, $m^2_\T$ from
\eq\nr{eff_params} may show temperature dependence.
This implies that the free energy density carried by the 
scalar field, 
$
 V \;\equiv\; { m^2_\T \varphi^2 } / {2}  
$, 
also contributes to the entropy density, 
$
 s = s^{ }_r - \partial^{ }_\T V
$, 
where $s^{ }_r$ is the contribution of thermal radiation. 
Physically, storing free energy in $\varphi$, whose value
carries definite information, decreases the total entropy.  
Then we should take care that the total entropy does not decrease
with time.  

Let us denote the partial derivatives of the  effective potential by  
\be
 \partial^{ }_\varphi V \;\equiv\;  
  m^2_\T \varphi
 \;, \quad
 \partial^{ }_\T V \;\equiv\; \frac{ 
 \bigl( \partial^{ }_\T m^2_\T \bigr) \varphi^2 }{2} 
 \;, \la{dTV}
\ee
and the energy density and  
pressure of the Yang-Mills plasma
by $e^{ }_r$ and $p^{ }_r$, respectively.
In equilibrium,  
the corresponding total variables are
$
 e = e^{ }_r + V - T \partial^{ }_\T V
$, 
$
 p = p^{ }_r - V 
$. 
The condition of entropy increase can now be imposed as 
\be
 T \partial^{ }_t \bigl[ 
 \bigl( s_r - \partial^{ }_\T V \bigr) a^3 \bigr] 
 = a^3 
 \Upsilon^{ }
 \dot{\varphi}^2(t)
 \;. \la{eom_sr}
\ee
Making use of 
$e^{ }_r =  T s^{ }_r - p^{ }_r$
and 
$s^{ }_r = \partial^{ }_\rmii{$T$} p^{ }_r$ , 
this can equivalently be rewritten as an equation for the energy density, 
\be
 \dot{e}^{ }_r + 3 H \bigl( e^{ }_r + p^{ }_r - T \partial^{ }_\T V \bigr)
 - T \partial^{ }_t \bigl( \partial^{ }_\T V \bigr)
 = 
 \Upsilon^{ }
 \dot{\varphi}^2(t)
 \;. \la{eom_er}
\ee
We note that \eq\nr{eom_er} can alternatively be obtained by writing 
the energy-momentum tensor as a sum of the radiation and
field parts, and imposing its overall conservation~\cite{hydro1}.

To close the system, the equations above are
supplemented by the Friedmann equation,  
\be
 H = \sqrt{\frac{8\pi}{3}} 
 \frac{\sqrt{ e^{ }_r + V -T \partial^{ }_\T V  
  + \dot{\varphi}^2/2 }}
      { \mpl }
 \;, \la{eom_H}
\ee
where we have assumed a spatially flat universe.  
Eqs.~\nr{eom_er}, \nr{eom_H} apply for any type of a radiation
equation of state, however for our numerical solution in
\se\ref{se:numerics} we adopt the simple conformal form 
\be 
  p^{ }_r =  \frac{g^{ }_* \pi^2 T^4 }{ 90 }
 \;, \quad
 T \gg \Lambdamsbar
 \;, \la{p_r}
\ee
with $g^{ }_*$ constant; 
$e^{ }_r$ and $s^{ }_r$ follow from this as stated above \eq\nr{eom_er}. 

If $\partial^{ }_\T m_\T^2 = 0$, 
\eqs\nr{eom_er} and \nr{eom_H} reduce to the standard form 
employed in refs.~\cite{warm1,warm2,warm3,warm4}. We also find
that the effects from $\partial^{ }_\T m_\T^2$ are numerically
small for the benchmarks of \se\ref{se:numerics}, 
as illustrated by the nearly constant values 
of $m^{ }_T$ in \fig\ref{fig:rates}.

%
\section{Retarded pseudoscalar correlator}
\la{se:C_R}

Having suggested a formalism depending on $C^{ }_\R$, 
the remaining step is to specify its form.
In this section we motivate a simple but semi-realistic 
approximation for $C^{ }_\R$. 

In a weakly coupled 
thermal system, $C^{ }_\R$ and the corresponding
spectral function, $\rho(\omega) = \im C^{ }_\R(\omega + i 0^+_{ })$, 
contain a lot of structure. At $\omega \gg \pi T$, $C^{ }_\R$ is
dominated by a vacuum part, 
up to power-suppressed corrections~\cite{sch}; 
at $\omega \sim \pi T$, it develops 
substantial thermal modifications~\cite{Bulk_wdep}; 
at $\omega \sim g T$, features originate from collective plasma
excitations and Debye screening~\cite{Bulk_wdep}. For very small frequencies, 
$\omega \ll \alpha^2 T$, yet another behaviour takes over, dominated 
by the non-perturbative dynamics of colour-magnetic fields~\cite{sphaleron}. 
A numerical evaluation of the contribution 
of the scales $\omega \sim g T$ and $\pi T$~\cite{Bulk_wdep}
suggests that it is overshadowed by
the contribution from very small frequencies. 
Therefore, we adopt a model in which $C^{ }_\R$ 
only contains a vacuum part from
$\omega \gg \pi T$ and an infrared part from $\omega \lsim \alpha^2 T$.
Furthermore, the ``tails'' of the two contributions 
are numerically small outside their domains of validity,
so we can establish
an interpolation simply by adding the parts together, 
\be
 C^{ }_\R(\omega) 
 \simeq 
 C^\rmi{vac}_\R(\omega)
 + 
 C^\rmiii{IR}_\R(\omega)
 \;. 
\ee

Starting with $C^\rmi{vac}_\R(\omega)$,
which is known up to NLO~\cite{old,Bulk_OPE} and even beyond it,  
we restrict here to the leading-order part, 
\be
 \frac{C^\rmi{vac}_\RB(\omega)}{16 \dA c_\chi^2 } 
 \; \approx \; 
 \frac{g^4 \omega^4 }{(4\pi)^2 f_a^2}
 \biggl\{ 
   \frac{1}{\epsilon} + 2 \ln \biggl( \frac{i\bmu}{\omega} \biggr) - 1 
 \biggr\} 
 \;, \quad
 \dA \; \equiv \; \Nc^2 - 1 
 \;, \quad
 c^{ }_{\chi} \; \equiv \; \frac{1}{64\pi^2}
 \;. \la{C_R_vac}
\ee 
Here $D = 4 - 2\epsilon$ is the space-time dimension,  
$g^2$ denotes the renormalized coupling, 
\be
 g^2(\bmu) \approx \frac{1}{2 b^{ }_0 \ln(\bmu/\Lambdamsbar)}
 \;, \quad
 b^{ }_0 \; \equiv \; \frac{11\Nc}{3(4\pi)^2}
 \;, \la{gw2}
\ee
and $\bmu$ is the renormalization scale of the $\msbar$ scheme. 

Before turning to further technical specifications, let us 
elaborate on the physical meaning of \eq\nr{C_R_vac}. Because of the 
logarithm, $C^{ }_\R$ has an imaginary part along the positive real 
axis ($\im \ln i = \pi/2$). 
This corresponds to the decay width of the scalar particle 
into gauge bosons ($\varphi\to gg$), and yields the vacuum 
contribution to $\Upsilon$, cf.\ \eq\nr{eff_params}. The real 
part of $C^{ }_\R$ yields a mass correction, cf.\ \eq\nr{eff_params}, 
but also includes a divergence, which necessitates the introduction
of a counterterm $\delta Z$ in \eq\nr{eom2}. Even though the presence
of a divergence may sound like a formal issue, it is closely
related to the existence of the logarithm, and can to some extent
be viewed as a consequence of the presence of a physical decay width. 

Returning to technical details, we note the presence of the scale 
parameter $\bmu$ in two different locations, \eqs\nr{C_R_vac} and \nr{gw2}. 
The scale parameter is just an auxiliary quantity, and must cancel from
physical results. However, the way in which this happens in the two
occurrences is conceptually different. The $1/\epsilon$ divergence
and $\ln\bmu$ in \eq\nr{C_R_vac} are reflections of the fact that 
\eq\nr{L} is non-renormalizable; the divergence can be cancelled
by a counterterm, denoted by $\delta Z$ 
in \eq\nr{eom2}. As a result 
the combination $1/\epsilon + \ln\bmu^2$ gets replaced
with a physical logarithm $\ln \Lambda^2$, where $\Lambda$
characterizes the UV completion of the theory. From the low-energy
perspective, $\Lambda$ is an additional free parameter. 
Its contribution is supposed to be numerically small
compared with the tree-level term $m^2$
within the domain of \eq\nr{consistency},  
and we set $\Lambda \equiv f^{ }_a$ from now on. 
We note that if $\omega\sim m \sim f^{ }_a / \alpha$, 
then $\re C^{ }_\R < 0$, yielding a positive mass correction
according to \eq\nr{eff_params}.

In contrast, the scale parameter appearing in \eq\nr{gw2} 
is related to renormalizable 
physics of the Yang-Mills plasma, and gets cancelled
by higher-order contributions. 
We can incorporate their main effects simply by choosing $\bmu$
according to the physical scales that appear, thereby
eliminating large logarithms.  
As a qualitatively reasonable recipe we set 
\be
 \bmu 
  \stackrel{\rmii{\nr{gw2}}}{=} 
 \sqrt{(2\pi \Lambdamsbar)^2 + (2\pi T)^2 + |\omega|^2}
 \;, \qquad
 \mbox{min}(2\pi T, |\omega|) \gg 2\pi \Lambdamsbar
 \;. 
 \la{scale}
\ee

Let us then turn to the IR part of $C^{ }_\R$. 
The benchmark information here is that around origin,  
the spectral function goes over into a ``transport coefficient'',
\be
 \lim_{\omega\to 0} \frac{2T\im C^{ }_\R (\omega)}{\omega}
 = 
 \frac{ \Gamma^{ }_\rmi{diff} }{f_a^2} 
 \;, \la{intercept}
\ee
where $\Gamma^{ }_\rmi{diff}$ refers to the Chern-Simons diffusion rate. 
The Chern-Simons diffusion rate dominates the result in 
the domain $\omega \ll \Delta$, where $\Delta$ is from 
\eq\nr{Delta_appro}. The determination of 
$\Gamma^{ }_\rmi{diff}$ requires a Monte Carlo simulation. 
For us it is 
sufficient to transcribe the numerical result from ref.~\cite{sphaleron}
into a form similar to that in \eq\nr{C_R_vac}, 
\be
 \im C^{ }_\R (\omega) 
 \stackrel{\omega \,\ll\, \Delta}{\equiv} 
 \Upsilon^{ }_{\rmiii{IR}}\, \omega
 \;, \quad 
 \frac{ \Upsilon^{ }_{\rmiii{IR}} }
      { 16 \dA c_\chi^2 } 
 \sim 
 \frac{1.2 g^4 (g^2 \Nc^{ } T)^3 }{16 \pi f_a^2}
 \biggl(
  \ln \frac{\mE^{ }}{\gamma}
  + 
 3.041 
 \biggr)
 \;,
 \la{rho_T_ir}
\ee
where $\gamma$ is a solution of the equation
\be
 \gamma = \frac{g^2 \Nc^{ }T}{4\pi}
 \biggl(
  \ln \frac{\mE^{ }}{\gamma}
  + 
 3.041 
 \biggr)
 \;, \la{gamma_eq}
\ee
and $\mE^{2} \equiv g^2 \Nc^{ }T^2/3$ 
is the Debye mass squared of Yang-Mills plasma.

Concerning the full $C^\rmiii{IR}_\R$, there is 
neither a numerical determination nor a good understanding
concerning its $\omega$-dependence. However, there is a belief that 
the Chern-Simons number should undergo diffusive motion; 
if so, it may be described by the Langevin equation, 
which in turn gives rise to a Lorentzian spectral shape
(cf.,\ e.g.,\ ref.~\cite{basics}).
Therefore we assume in the following that  
\be
 C^\rmiii{IR}_\R(\omega) 
 \;\stackrel{\omega\sim\Delta}{\simeq}\;
 -\frac{\omega\Delta \Upsilon^{ }_{\rmiii{IR}}}{\omega + i \Delta}
 \;. \la{CR_T_ir}
\ee
Here $ \im C^\rmiii{IR}_\R $ is chosen to match 
\eq\nr{rho_T_ir} for $|\omega| \ll \Delta$, and 
the constant part of $\re C^\rmiii{IR}_\R$ is chosen 
so that $C^\rmiii{IR}_\R(0) = 0$,
in accordance with the discussion below \eq\nr{L}.
For $\omega\neq 0$, 
\eq\nr{CR_T_ir} yields $\re C^{ }_\R < 0$, 
corresponding to a positive mass correction
according to \eq\nr{eff_params}.

The value of $\Delta$ is unknown, so we assume
\be
 \Delta = c \,
 \biggl( \frac{g^2\Nc}{4\pi} \biggr)^2_{ } T
 \;, 
 \quad
 c \simeq 10 
 \;.
 \la{Delta}
\ee
The largish coefficient is chosen in order to make 
\eq\nr{CR_T_ir} as flat as possible, being therefore  
maximally welcoming to the approximation 
adopted in refs.~\cite{warm1,warm2,warm3,warm4}.

%
\section{Examples of numerical solutions}
\la{se:numerics}

Given $C^{ }_\R$ from \se\ref{se:C_R}, 
we can solve for $\Upsilon$ and $m_\T^2$ from \eq\nr{eff_params}. 
Subsequently \eqs\nr{eomF5}, \nr{eom_er} and \nr{eom_H} can be 
integrated with given initial conditions. 

For representing the medium, 
we adopt the QCD-like values 
$\Nc^{ }=3$, $g^{ }_* = 2 \dA \stackrel{\rmiii{axion}}{+} 1= 17$, 
$\Lambdamsbar = 0.2$~GeV.
We stress, however, that this serves purposes
of illustration only: we do not otherwise constrain the parameters
through QCD-like axion physics. In the real world, the
gauge group could rather be a unified one, and the scale 
parameter could be different. 
However, a small $\Lambdamsbar$ simplifies the analysis, 
because then $\alpha$ is small (numerically $\alpha\sim 0.015$) 
and we can use a conformal equation of state for thermal radiation
in \eq\nr{p_r}, as long as the initial temperature 
is $\gg \Lambdamsbar$, e.g.\ $T(0) = 10^{-10}\,\mpl$.

The key parameters affecting the solution are 
$m$, $f^{ }_a$, and $\varphi(0)$.\footnote{%
  The initial 
  $\dot\varphi$ is unimportant as the slow-roll
  solution is an attractor; we take
  $\dot\varphi(0) \simeq - m^2 \varphi(0) / [3 H(0)]$. } 
We have adopted two benchmarks to illustrate the dynamics, 
one leading to the weak ($Q\lsim 1$)
and the other to the strong regime ($Q \gg 1$) of warm inflation: 
\ba
 && m = 7 \times 10^{-7} \mpl \;, \quad
 f^{ }_a = 8 \times 10^{-7} \mpl \;, \quad
 \varphi(0) = 4 \, \mpl
 \;, \qquad (\mbox{weak~regime}) 
 \la{benchmark_weak}
 \\ 
 && m = 7 \times 10^{-7} \mpl \;, \quad
 f^{ }_a = 2\times 10^{-7} \mpl \;, \quad
 \varphi(0) = 2 \, \mpl
 \;. \qquad (\mbox{strong~regime}) \hspace*{6mm}
 \la{benchmark_strong}
\ea
In both cases \eq\nr{consistency} is rather marginally satisfied, 
and this turns out to be essential for the dynamics: 
if $f^{ }_a$ is further decreased, the vacuum part of $C^{ }_\R$
starts to dominate but the framework becomes 
theoretically inconsistent; 
if $f^{ }_a$ is increased, the effects from $C^{ }_\R$
disappear, and we return to usual (cold) chaotic inflation.

\begin{figure}[t]

\hspace*{-0.1cm}
\centerline{%
  \epsfxsize=7.5cm\epsfbox{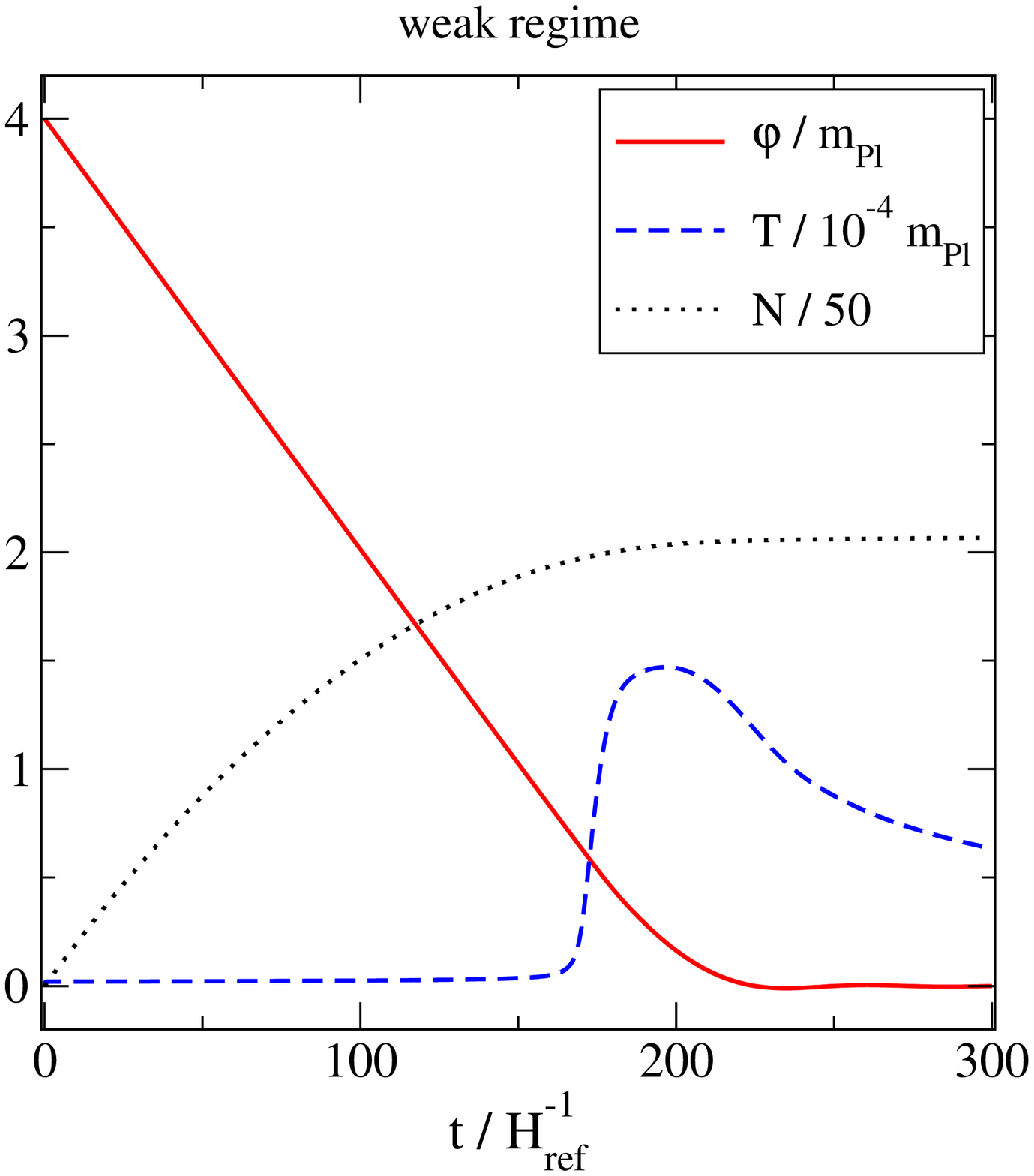}
~~\epsfxsize=7.5cm\epsfbox{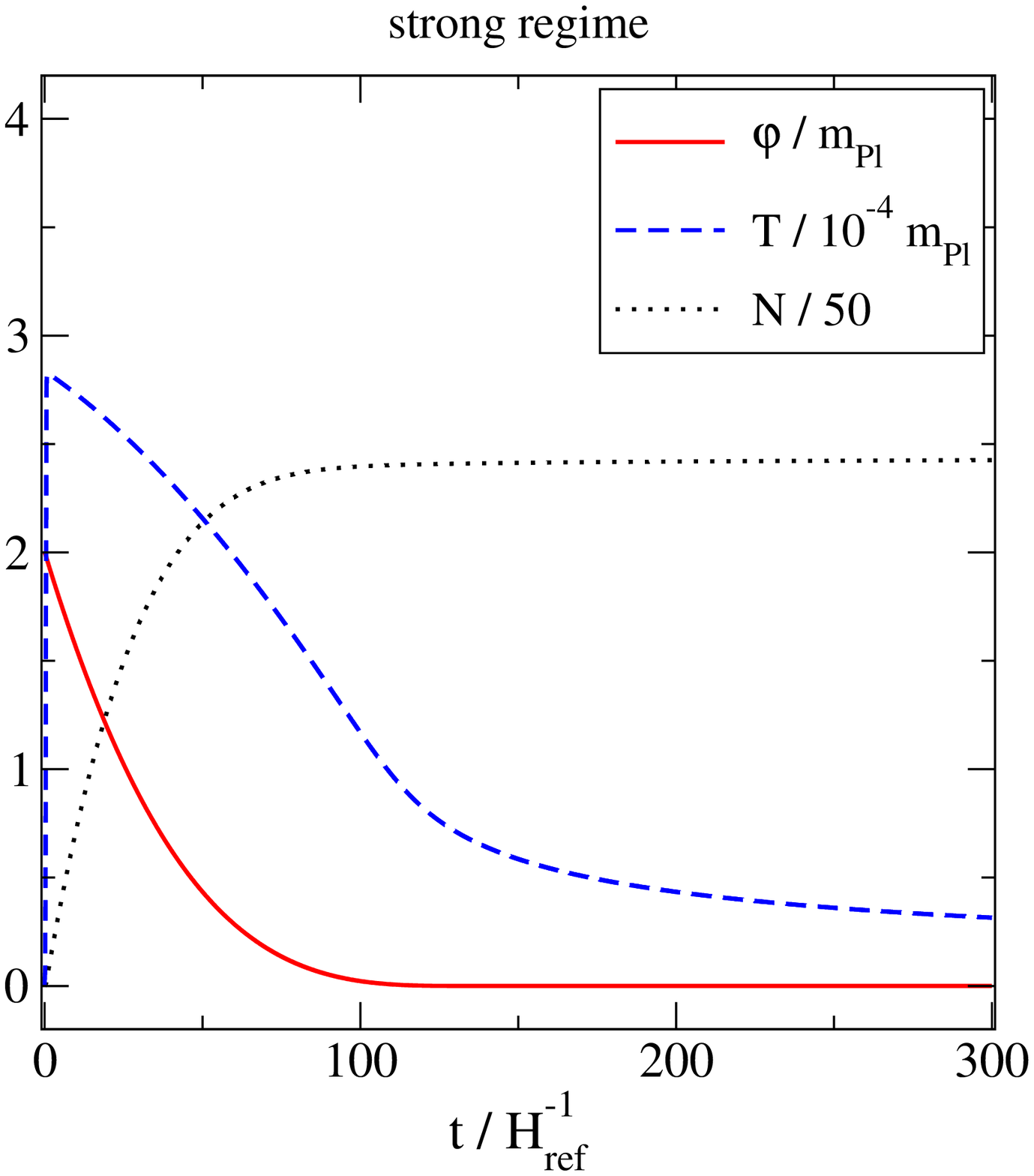}
}

\caption[a]{\small
 Left: the variables 
 $
  \varphi / \mpl
 $, 
 $ 
  T / ( 10^{-4} \, \mpl  )
 $, and 
 $
  N / 50 
 $ (number of $e$-folds), 
 for the parameters 
 in \eq\nr{benchmark_weak}, 
 as a function of $t / H^{-1}_\rmi{ref}$, 
 where $H^{ }_\rmi{ref}$ is from \eq\nr{Href}. 
 Right: the same for 
 \eq\nr{benchmark_strong}.
}

\la{fig:variables}
\end{figure}

As far as phenomenology goes, the amplitude of scalar perturbations, 
$A^{ }_s$, can always be chosen to match the observed value, 
by tuning $m/\mpl$. Currently the most stringent test 
comes from whether the spectral tilt, $n^{ }_s$, matches 
the Planck result~\cite{planck}.\footnote{%
  As far as other predictions go, 
  the tensor-to-scalar ratio $r$ is argued
  to be small in warm inflation,
  as scalar perturbations are increased
  by thermal fluctuations but tensor perturbations supposedly not, 
  even if we note that thermal fluctuations of
  a weakly coupled scalar field do
  yield a substantial contribution to the gravitational wave
  production rate~\cite{fluct}.  
  Non-Gaussianities are argued to offer for 
  a characteristic signature of warm inflation~\cite{warm_pred_2}. 
  However, for both of these observables 
  measurements only give upper bounds for now.  
 }  
A challenge here is that as the 
solution may interpolate between the weak and strong regimes, the
corresponding predictions need to be adopted from numerical 
work, which is typically specific to a particular 
model or parametric form of $\Upsilon$
(cf.,\ e.g., 
refs.~\cite{warm_pred_1,warm_pred_15,warm_pred_16,warm_pred_3}
and references therein).
In any case,  according to ref.~\cite{warm4}, the weak regime
could lead to a phenomenologically viable value of $n^{ }_s$, 
whereas ref.~\cite{warm1} found that the strong regime only works
by adding a constant to the potential, i.e.\ by considering 
hybrid rather than chaotic inflation. Here we have added no 
constant, and suspect that our benchmark points do not produce 
the correct $n^{ }_s$. Nevertheless, we hope that they  
can clearly illustrate general features of the dynamics. 

\begin{figure}[t]

\hspace*{-0.1cm}
\centerline{%
  \epsfxsize=7.5cm\epsfbox{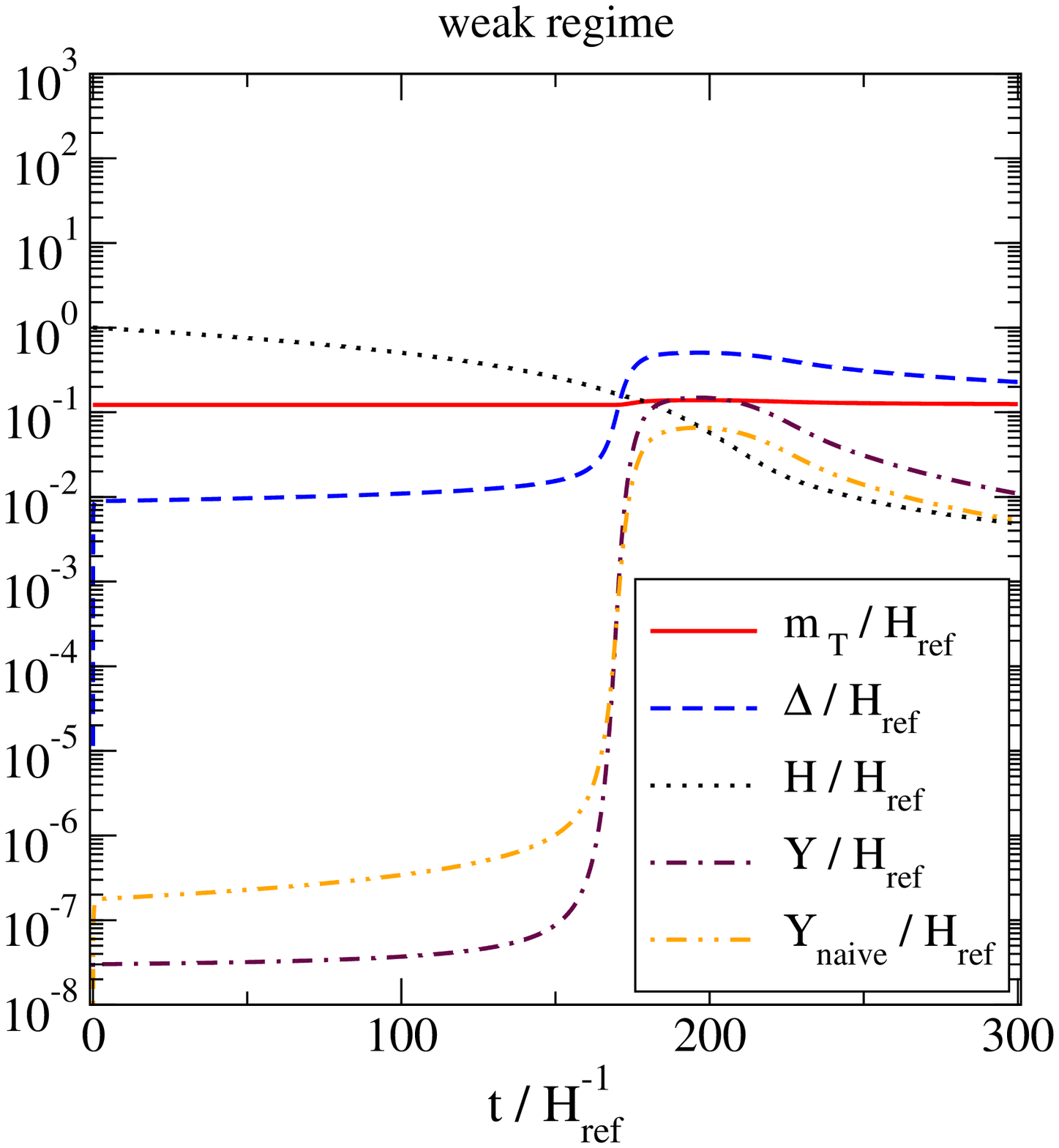}
~~\epsfxsize=7.5cm\epsfbox{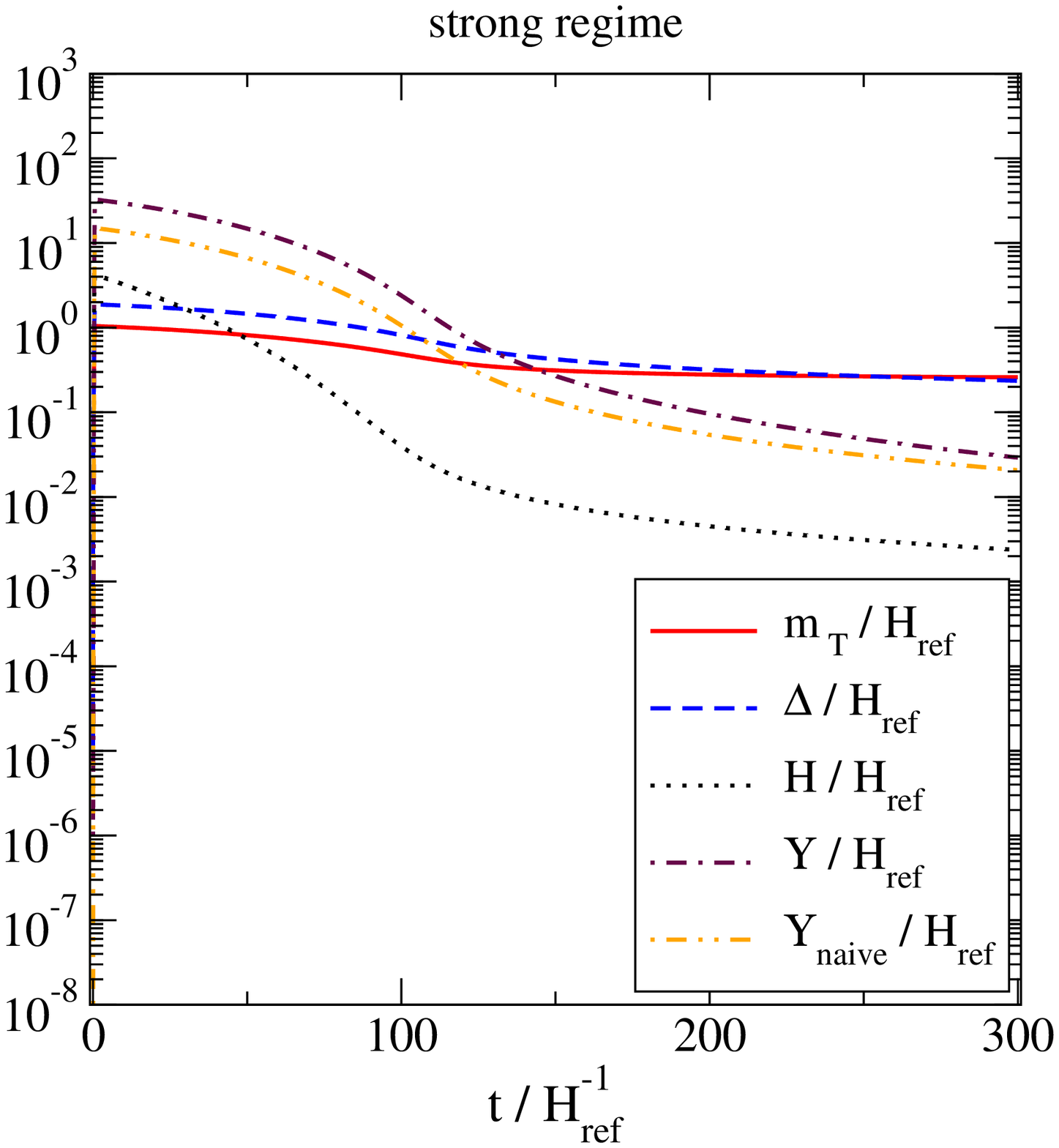}
}

\caption[a]{\small
 Left: the frequency scales 
 $
  m^{ }_\T 
 $, 
 $ 
  \Delta 
 $
 (cf.\ \eqs\nr{eff_params} and \nr{Delta})
 and the rates
 $
  H
 $, 
 $
  \Upsilon 
 $, 
 and 
 $ 
  \Upsilon^{ }_{\!\rmi{naive}} 
 $  
 for the parameters 
 in \eq\nr{benchmark_weak},
 normalized to $H^{ }_\rmi{ref}$ from \eq\nr{Href}. 
 Here $\Upsilon^{ }_{\!\rmi{naive}} \;\equiv\; \kappa\alpha^5 T^3 / f_a^2 $
 is the $\omega$-independent
 thermal width employed in refs.~\cite{warm1,warm2,warm3,warm4},
 shown for illustration but not affecting our dynamics 
 (we have inserted $\kappa = 10^2$ and $\alpha = 0.015$). 
 Right: the same for the parameters 
 in \eq\nr{benchmark_strong}.
 The conclusions drawn from these plots are 
 discussed around the end of \se\ref{se:numerics}. 
}

\la{fig:rates}
\end{figure}

Our numerical results for $\varphi$, $T$ and the number
of $e$-folds $N$ are plotted 
in \fig\ref{fig:variables}. 
As outlined in \se\ref{se:scales}, 
the key rates and frequency scales 
governing the nature of the solution are 
$\Upsilon$, $H$, $m^{ }_\T$ and $\Delta$, and they are 
plotted in \fig\ref{fig:rates}. 
These are conveniently scaled to 
the (approximate) initial Hubble rate, 
\be
 H^{ }_\rmi{ref} 
 \; \equiv \; 
 \sqrt{\frac{4\pi}{3}} \frac{m\, \varphi(0)}{ \mpl }
 \;. \la{Href}
\ee

We observe from \fig\ref{fig:rates}(left) that at the initial stage of a 
weak-regime solution, $\Delta \ll m^{ }_\T$. Then the vacuum term
in \eq\nr{Ups_vac} is the dominant component of $\Upsilon$. At a certain
moment, the temperature increases and the system rapidly moves 
into a domain in which $\Delta > m^{ }_\T$, so that \eq\nr{Ups_T}
dominates. Simultaneously, $\Upsilon$ becomes larger than the Hubble rate. 

Decreasing $f^{ }_a$ moderately, according to \eq\nr{benchmark_strong}, 
boosts both the vacuum and thermal $\Upsilon$ by an order of magnitude.
Then the temperature reaches its maximal value already at the beginning,
as shown in \fig\ref{fig:variables}(right).
Now $ \Delta > m^{ }_\T $ and $\Upsilon \gg H$ according
to \fig\ref{fig:rates}(right), and the dynamics
takes place in the strong regime. However, this is only 
achieved if the vacuum part of $\Upsilon$ (cf.\ \eq\nr{Ups_vac})
is sufficiently large at the beginning.  

%
\section{Conclusions and outlook}
\la{se:concl}

There has been renewed interest in models of warm axion inflation, 
as it was noted that adopting a more realistic thermal friction coefficient
than in early works changes the nature of the solution, 
possibly rendering a phenomenologically viable scenario and simultaneously 
addressing theoretical issues 
such as the swampland problem~\cite{warm1,warm2,warm3,warm4}. 
However, as duly acknowledged in these works, 
the new friction coefficient was still not fully realistic, 
as it was specific to a certain frequency domain, which may or may not 
be realized by the actual solution. The purpose of this paper
has been to present a theoretical framework which permits to eliminate
this approximation. 

Numerical benchmark results from our framework are shown in 
\figs\ref{fig:variables} and \ref{fig:rates}. We find
that reaching the strong regime of warm inflation requires fine-tuning,
as the vacuum contribution from the non-renormalizable operator, 
\eq\nr{Ups_vac}, needs to be substantial at early times, yet not so large
that it would render the framework inconsistent. A weak regime exists more 
broadly, as increasing $f^{ }_a$ decreases the significance of 
the non-renormalizable operator and connects us to 
the case of cold inflation. On the qualitative level, 
we confirm the existence of scenarios as proposed in 
refs.~\cite{warm1,warm2,warm3,warm4}, even if we find strong
dependence on the vacuum contribution in 
\eq\nr{Ups_vac}, which was omitted in those works. 
As a consequence, the  range of admissible
values of $f^{ }_a$ is more tightly constrained. 
We stress again 
that \eq\nr{Ups_T} employed in refs.~\cite{warm1,warm2,warm3,warm4}
is not active around the beginning, when $m \gg \alpha^2 T$. 

There are a number of directions in which our work could be expanded. 
One concerns the retarded pseudoscalar correlator $C^{ }_\R$, 
which determines the value of the vacuum or thermal friction coefficient. 
Here we adopted a reasonable model, but more information and NLO
corrections could be added in certain frequency domains~\cite{Bulk_wdep}. 
One obstacle is, however, that the infrared part of $C^{ }_\R$ is
currently poorly understood, with only the transport coefficient
in \eq\nr{intercept} estimated, but the frequency dependence subject
to argumentation (cf.\ \se\ref{se:C_R}). 
Knowing the frequency dependence would be essential, as it determines
the parameter $\Delta$ that dictates whether the vacuum
or thermal value of $\Upsilon$ should be used. 

Another extension would be to carry out parameter scans with
various potentials, mapping the possibly 
fine-tuned domains that may be phenomenologically
most viable. In principle there is no obstacle to 
attacking this challenge, apart from the fact that 
predictions rely on an interpolating function for the spectrum
of scalar fluctuations, 
whose determination involves model dependences. 
Before embarking on scans,
this function might deserve an independent re-evaluation
(cf.,\ e.g.,\ ref.~\cite{warm_pred_4}). 

\vspace*{-4mm}

%
\section*{Acknowledgements}

We thank Rudnei Ramos for helpful correspondence.
This work was partly supported by the Swiss National Science Foundation
(SNF) under grant 200020B-188712.

%
\appendix
\renewcommand{\thesection}{Appendix~\Alph{section}}
\renewcommand{\thesubsection}{\Alph{section}.\arabic{subsection}}
\renewcommand{\theequation}{\Alph{section}.\arabic{equation}}

\small{
%

}

\end{document}